# A Review of Micromachined Thermal Accelerometers

Rahul Mukherjee*, Joydeep Basu, Pradip Mandal, Prasanta Kumar Guha
Dept. of Electronics & Electrical Communication Engineering
Indian Institute of Technology Kharagpur, Kharagpur 721302, India
*E-mail: rahul10.iitkgp@gmail.com

***Abstract:*** Thermal convection based micro-electromechanical accelerometer is a relatively new kind of acceleration sensor that does not require a solid proof mass, yielding unique benefits like high shock survival rating, low production cost, and integrability with CMOS integrated circuit technology. This article provides a comprehensive survey of the research, development, and current trends in the field of thermal acceleration sensors, with detailed enumeration on the theory, operation, modeling, and numerical simulation of such devices. Different reported varieties and structures of thermal accelerometers have been reviewed highlighting key design, implementation, and performance aspects. Materials and technologies used for fabrication of such sensors have also been discussed. Further, the advantages and challenges for thermal accelerometers vis-à-vis other prominent accelerometer types have been presented, followed by an overview of associated signal conditioning circuitry and potential applications.

**1. Introduction:** Accelerometers have emerged as a ubiquitous sensor in recent times having high demand in the fields of consumer electronics, automotive, biomedical, defense, aerospace, navigation, and industrial applications [1]–[3]. The rapid progress of semiconductor fabrication technology has led to the development of predominantly silicon based micro-electromechanical system (MEMS) accelerometers that are gaining popularity due to features of small size, low power requirement, high performance, and low cost [2]. The physical acceleration required to be measured might be static, like gravitational acceleration; or dynamic, like vibration and shock. Some of the common sensor types include capacitive ([4], [5]), piezoelectric ([6], [7]), piezoresistive ([8], [9]), and tunneling accelerometers [10] – most of them have a solid proof mass which changes its position or shape due to the applied acceleration. Owing to the mechanical movement involved, such devices have lower shock survival rating along with other issue like stiction, mechanical ringing and hysteresis. In contrast, thermal accelerometers sense acceleration by measuring the displacement of a tiny heated fluid bubble present within a sealed cavity. As there is no solid seismic mass, the shock survivability of a thermal accelerometer is high. Its fabrication is also much simpler and the fabrication cost is low. Moreover, the integration of the sensor with complementary metal-oxide semiconductor (CMOS) signal conditioning integrated circuit (IC) on the same silicon die is also convenient.

Micromachined accelerometer based on convective heat transfer was first demonstrated in 1997 by Leung et al. [11]. Since then, this field has received appreciable research attention leading to reported devices with a number of performance aspects surpassing that of the counterparts. Nonetheless, the relative newness of the domain of MEMS thermal acceleration sensors coupled with the ever increasing scope of human-machine interaction offers ample scope for future research and innovation. This review article is meant to be an enabling medium in this direction, providing detailed insight on the operation, modeling, and performance aspects (like sensitivity, power and bandwidth) of thermal convective accelerometers. It surveys different reported varieties of thermal accelerometers, along with the materials and methods used by various research groups for fabrication of such accelerometers. A brief description of the finite element simulation of such sensor design and performance have been provided using a standard software package. Furthermore, commercial developments and potential applications of thermal accelerometers and its comparative evaluation with the dominating capacitive accelerometer type have been summarized. The prospects of integration of thermal inertial sensors with the necessary signal conditioning (amplification, filtering, etc.) electronics have also been surveyed, followed by a concluding discussion on the future research directions in this field.

**2. Working Principle:** The operation of thermal inertial sensors is based on the natural convection of fluid. A general structure of a single-axis thermal accelerometer is illustrated in figure 1(a) which consists of a micro cavity created by front-side bulk micromachining of silicon wafer. An electrical resistive heater is suspended at the center of the cavity and a pair of temperature sensors (like thermistor; or thermopile made of serially connected thermocouples) are placed symmetrically around the heater. The fluid (say air) present in the cavity remains encapsulated by an outer cover (package). Due to the heat dissipation of the heater, a hot thermal bubble of the fluid is formed surrounding it. In steady state (i.e., without any acceleration), the temperature profile within the cavity remains symmetrical with respect to the heater, and the symmetrically placed temperature sensors detect identical temperatures. However, in the presence of an applied acceleration, the temperature profile gets skewed due to physical displacement of the thermal bubble as shown in figure 1(b). The temperature increases on one side of the heater and decreases at the other side as shown in the profile in figure 1(c). The resultant differential temperature ($\Delta T$) is proportional to the applied acceleration and is measured by the temperature sensors.

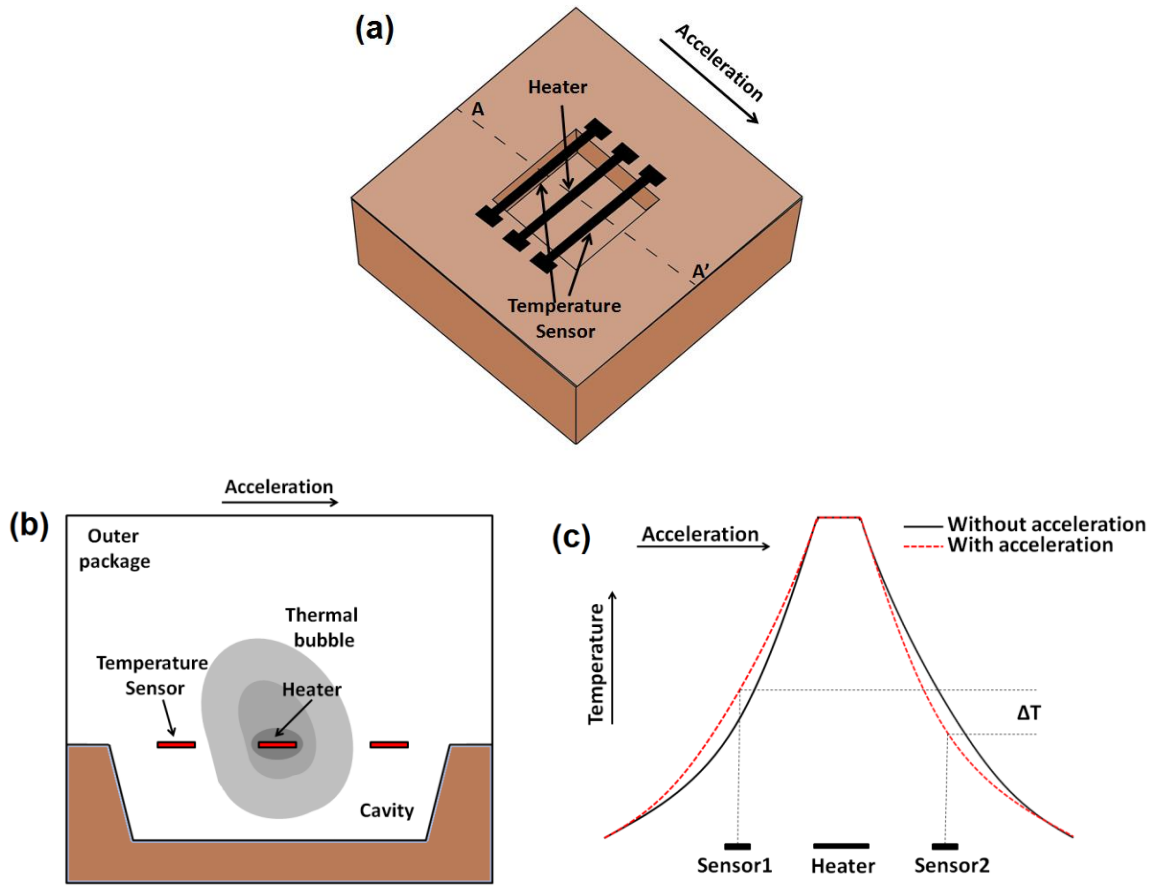

Figure 1. (a) Schematic view of thermal accelerometer, (b) cross-sectional view along AA′ line, and (c) temperature profile along AA′.

**3. Modeling and Simulation:** In order to predict the performance of a particular thermal inertial accelerometer design, or equivalently, to be able to tune the physical design to obtain the desirable performance, description of such devices via mathematical modeling is imperative.

**3.1: Analytical Modeling:** The device performance can be mathematically analyzed by modeling the accelerometer using simple geometries. For the structure in figure 1(a), which has a centrally placed heater and working fluid enclosed by cavity and outer cover, the heater can be modeled as a cylindrical heat source and the outer cover as a larger cylinder at ambient temperature [12]. The model can be even simplified to a spherical structure where the heater is a spherical source, and the outer cover is a larger sphere centering the heater and kept at ambient temperature [13], [14]. Such a simplified geometry is shown in figure 2(a). Here, $r$ and $\theta$ are the radial distance and angle of the spherical coordinate system respectively. The inner sphere represents the heater with radius $r_i$ and surface temperature $T_i$. On the other hand, the outer sphere represents the cavity wall surface having radius $r_o$ and wall temperature $T_o$ (where $T_i > T_o$). AA' represents the vertical axis along midsection. Ratio of the outer to inner sphere radius is $R$ (= $r_o/r_i$). The concentric cylinder model is generally used for single-axis accelerometers, while concentric sphere for the dual axis ones. In three-axis accelerometers, the x and y-axes accelerations are applied in-plane while the z-axis acceleration is applied out-of-plane and these too can be modeled with concentric spheres.

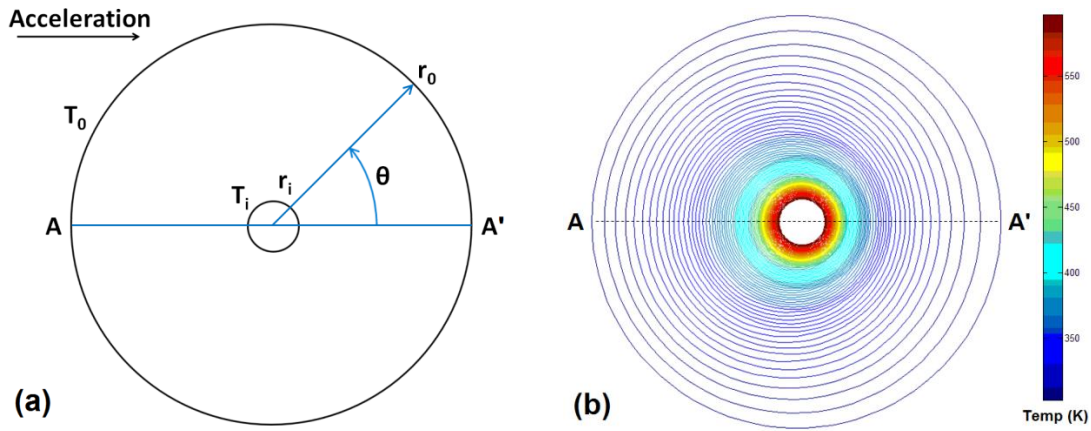

Figure 2. (a) Simplified model of thermal accelerometer represented using concentric spheres. (b) Temperature contour plotted by solving the governing equations.

The governing equations predicting temperature profile of a thermal accelerometer device are based on the principle of conservation of mass, momentum and energy [15] which are as follows:

$$\frac{\partial \rho}{\partial t} + \nabla.(\rho \mathbf{u}) = 0 \qquad (1)$$

$$\rho \left( \frac{\partial \mathbf{u}}{\partial t} + \mathbf{u}.\nabla \mathbf{u} \right) = -\nabla p + \nabla.\mathbf{I} + \mathbf{f} \qquad (2)$$

$$\rho C_p \left( \frac{\partial T}{\partial t} + u.\nabla T \right) = k\nabla^2 T \qquad\qquad (3)$$

Here, **u** is the flow velocity vector field, $\nabla$ is spatial divergence operator, $p$ is the pressure, **I** is the total stress tensor, **f** denotes the body forces acting on the fluid. The parameters $C_p$, $\rho$ and k are the specific heat, density and thermal conductivity of the fluid in the cavity, respectively. In the realm of micro-fluidics, various parameters determine the convective and conductive thermal energy flow in a fluid. For simplification, these parameters are clubbed together to define some non-dimensional numbers as stated below. Use of these dimensionless numbers help to simplify the governing equations as well as the overall analysis.

(i) *Fourier number* is defined as the ratio of the rate of heat conducted through a body to the rate of heat stored.

$$\text{Fourier number: } F_0 = \frac{\tau\alpha}{r_i^2} \qquad\qquad (4)$$

where, $\tau$ is the characteristic time, and $\alpha$ is the thermal diffusivity of the fluid defined as the ratio of heat conducted to heat stored which essentially represents how fast the heat diffuses through a material and is expressed as:

$$\alpha = \frac{k}{\rho C_p} \qquad\qquad (5)$$

Higher value of the Fourier number indicates a faster heat propagation through the body.

(ii) *Prandtl number* represents the ratio of diffusion of momentum to diffusion of heat in a fluid.

$$\text{Prandtl number: } Pr = \frac{\mu C_p}{k} \qquad\qquad (6)$$

Here $\mu$ is the dynamic viscosity of the fluid. For air, the Pandtl number is around 0.7–0.8 and the value is higher for oils (e.g., SAE10, SAE40 etc.). The sensitivity of thermal accelerometer, depends upon the Prandtl number of the fluid present in the cavity. With higher values of Prandtl number of the working fluid, the sensitivity of the device increases.

(iii) *Grashof number* is the ratio of buoyancy force to viscous force.

Grashof number: $Gr = \dfrac{\rho^2 a \beta (T-T_0)}{\mu^2} l^3$           (7)

where, $a$ is the applied acceleration, $\beta$ is coefficient of fluid expansion, $l$ is the characteristic length of the device and $(T-T_0)$ is the temperature difference between the heater and the bulk substrate. As the sensitivity of convective accelerometer depends on the working fluid, characteristic length of the device and the temperature difference between the heater and the bulk substrate, the Grashof number qualitatively indicates the sensitivity of the device.

(iv)  *Rayleigh number* is the product of Prandtl and Grashof numbers.

Rayleigh number:  $Ra = Pr.Gr$           (8)

The sensitivity of thermal accelerometer is directly proportional to the Rayleigh number [16].

The governing equations (1)–(3) can be solved using the boundary conditions of the thermal accelerometer under consideration to obtain the temperature distribution and the velocity profile. The temperature distribution ($T'$) can be obtained as follows [17], [18]:

$T' = T_0 + (Gr.Pr)T_1 + \left(Gr.Pr\right)^2 T_2 + \left(Gr.Pr\right)^3 T_3 + ...$           (9)

Where, $T_0, T_1, T_2, T_3, \ldots$ are dimensionless functions of radius ratio ($R$) and the radial distance ($r$). This can be further approximated as:

$T' = T_0 + (Gr.Pr)T_1$           (10)

Here, $T_0$ is the temperature profile due to thermal conduction from the inner to the outer sphere, and $T_1$ is the temperature profile due to fluid convection, as given by:

$T_0 = -\dfrac{1}{R-1} + \dfrac{R}{R-1} \cdot \dfrac{1}{r}$           (11)

$T_1 = \left( C_1 R^2 + A_3 R + C_2 + C_3 \dfrac{1}{R} + A_4 \dfrac{1}{R^2} + C_4 \dfrac{1}{R^3} + C_4 R \ln R \right) \cos\theta$           (12)

where, $C_i$ and $A_i$ are coefficients which are functions of R. It is to be noted that the governing equations can also be solved using numerical computing tools like MATLAB [19] to obtain the temperature contours (e.g., in figure 2(b)) and temperature profile inside the cavity and, hence the device sensitivity.

### 3.2: Numerical Simulation:

As the structure of a practical accelerometer is quite complex, it cannot be easily analyzed by means of the analytical equations. To study its various performance parameters like temperature profile etc., numerical simulators are required to be used. Different software tools are commercially available to this end, like ANSYS [20], CoventorWare [21], and COMSOL Multiphysics [22]. Here, an example has been presented in COMSOL to illustrate the numerical simulation steps involved in Finite Element Method (FEM) analysis of thermal accelerometers. The tool provides an environment where effects of different intercoupled physical phenomena can be simulated together to predict the overall effect. Through a number of *physics user interfaces*, information corresponding to the physical phenomena can be incorporated by means of variables and equations. For a convective accelerometer, to get the effect of acceleration on the heated fluid within the cavity, *Joule heating* physics and *laminar flow* physics can be used in the numerical simulator [22]. The devices may be specified in 2D (two dimensions) or in 3D (three dimensions). Compared to the 2D modeling where the simulator considers dimension in one axis (z-axis) is infinite, in 3D, all dimensions are finite. As expected, simulation in 3D provides more accurate results but at the cost of CPU time.

Figure 3 illustrates an example of a micromachined dual-axis convective accelerometer having a planar square-shaped heater placed at the center of a cuboid cavity. Using the same principle as a single-axis thermal accelerometer as discussed in Sec. 2, a dual-axis accelerometer measures acceleration along two orthogonal axes using two pairs of temperature sensors placed equidistant from the central heater. The heater plate is supported by four arms clamped from the four corners of the cavity. The arms may be made of polysilicon or metal (as electrical conductor), and oxide and/or nitride layers (as insulating layer around the conductor). For the purpose of finding the temperature distribution due to the heater within the cavity, the temperature sensing structures may be omitted. The working fluid (here air) also needs to be defined with its properties set as functions of temperature. In this example, the temperature of the outer walls (cavity and top cover) of the device has been set at 300K. The device has been meshed in *free tetrahedral* mode [22], [23]. The critical regions like the square heater corner regions and tapered regions of the supporting arms have been meshed with higher resolution (smaller size of elements) to get better accuracy. The heat flows from the heater to the surrounding air as well as into the supporting arms creating the temperature contour as seen in figure 3(a). Further, acceleration has been applied towards the right side using *volume force* of *laminar flow* physics in the air volume. This produces the skewed temperature contour of figure 3(b) and (c). The temperature at any point of the structure, or along any straight line can be plotted using *stationary study*. With the help of these data, the sensitivity of the device can be determined by subtracting the temperature at two opposite equidistant points from the heater (prospective location for the temperature sensors). With the help of *time dependent* study in simulation, the transient response and hence, the -3dB bandwidth of the sensor can also be determined [24], [25]. One can play with the heater and cavity shape/size, constituent material and fluid properties, etc. for improving the sensitivity of the device. Apart from sensitivity and bandwidth, other relevant performance parameters (specifications) are its measurement range, linearity, resolution, voltage noise density or noise equivalent acceleration (NEA), overload shock limit, power consumption, size etc. [26], [27].

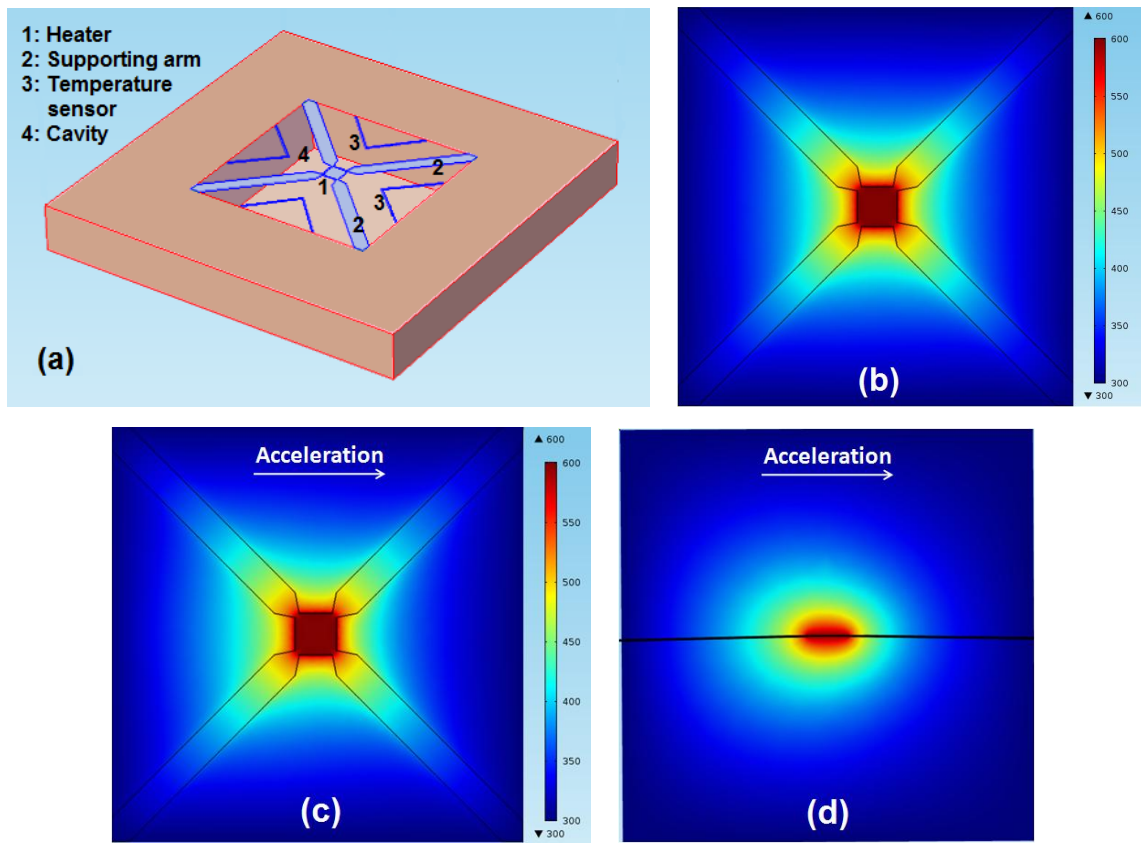

Figure 3. (a) Dual-axis convective accelerometer with square plate-shaped heater with supporting arms and temperature sensing structures. Temperature contour plots due to the heater: (b) Top view without any acceleration; (c) Top view and (d) side view with 100g acceleration (g is the acceleration due to gravity).

**Table 1.** Summary of reported performance of a few thermal convective accelerometers.

| Type, Technology | Heater material, Shape | Temp. sensor type, Shape | Working fluid | Sensitivity | Bandwidth/ Response time | Meas./ linearity range | Resolution/ RMS-noise | Power/ Temp. | Year & Ref. |
|---|---|---|---|---|---|---|---|---|---|
| Single-axis, Si MEMS | Poly-Si, bridge | Poly-Si thermistor, bridge | Air | 60mV/g | 20Hz | ±1g | 0.5mg | 20mW | 1997, [11] |
| Single-axis Si, MEMS | Pt, bridge | Pt thermistor, bridge | Air | 1bar: 2.5mV/g 25bar: 138mV/g | 20Hz | 3g | 0.3mg | 54mW | 2003, [36] |
| Single-axis Si, MEMS | Pt, bridge | Pt thermistor, bridge | Air | 0.12°C/g | 120Hz | ±2g | 0.25mg | 70mW | 2008, [24] |
| Single-axis, Si MEMS | Pt, bridge | Pt thermistor, bridge | He | 2.15bar: 0.002°C/g | 320Hz | ±2g | — | 300°C | 2011, [42] |
| Single-axis, Si MEMS | Pt, bridge | Pt thermistor, bridge | — | 0.0045°C/g | — | 10,000g | — | 200°C | 2011, [39] |
| Dual-axis, Si MEMS | Poly-Si, diamond | Al/Poly-Si thermopile | $SF_6$ | 3.5mV/g | 25Hz | 5g | — | 7.4mW | 2011, [16] |
| Dual-axis, SOI MEMS | Si, circular | Si thermistor, ring-shaped | Air | 13mV/g | 250Hz | ±5g | 10mg | 12.5mW | 2007, [35] |
| Tri-axis, MEMS+ Polymeric | Al on polyimide membrane | Al thermistor on polyimide membrane | Air | X, Y: 8mV/g Z: 2.2mV/g | 4Hz | ~0.6g | — | 45mW | 2011, [47] |
| Single-axis, CMOS MEMS | Poly-Si, bridge | Poly-Si thermistor, bridge | Air | 375mV/g | 14.5Hz | 10g | 30mg | 35mW | 2006, [25] |
| Single-axis, CMOS MEMS | Pt, bridge | Pt thermistor, bridge | $N_2$ | 0.034°C/g | 1025Hz (closed loop) | — | — | 70mW | 2012, [90] |
| Dual-axis, CMOS MEMS | Poly-Si, meander | Al/ Poly-Si thermopile | Air | 0.024°C/g | — | 150g | — | 200°C | 2010, [60] |
| Tri-axis, CMOS MEMS | Poly-Si, square | Poly-Si thermistor | Air | X: 8.8mV/g Y: 12.6mV/g Z: 0.45mV/g | 20Hz | 3g | X, Y: 2.6mg Z: 60mg | 10mW | 2014, [62], [63] |

**4. Varieties of Thermal Accelerometers:** A selection of unique and representative reports of micromachined thermal accelerometers have been reviewed here for providing the reader with an insight of the different design aspects that might influence the performance of such devices. Also, table 1 provides an outline of the performance of some of the reported accelerometers.

**4.1: Thermal Accelerometer in MEMS Process:**

The concept of thermal convective accelerometer without solid proof mass was first patented in 1996 by Dao et al. [28]. Subsequently, Leung's group from Simon Fraser University was the first to implement such accelerometers by custom fabrication on silicon substrate [11], [29]. Although the implemented devices were single-axis ones, but, dual-axis thermal accelerometers were also conceptualized. A cavity of dimension 1.5mm×4mm was generated by front-side bulk micromachining of Si wafer. Lightly doped polysilicon was used to realize the heater and thermistor bridges (1.5µm thick, 10µm wide and 1500µm long). Poly-Si heater is quite popular because of its higher sheet resistance compared to metals, and thus, suitable resistance can be achieved in a miniaturized portion. However, polysilicon shows long-term drift of electrical resistance owing to electro-migration [30], [31]. The convection accelerometer device was sealed in ceramic DIP-16 (dual inline package). The output sensitivity was high (~60mV/g at a heater power of 20mW) due to a large cavity size. It was shown experimentally that the device sensitivity increases linearly with the heater power and the square of the air pressure within the cavity. The measured frequency response was from DC to 20Hz. A relatively lower bandwidth is characteristic of thermal accelerometers (compared to capacitive accelerometers) owing to the slower response of thermal exchange.

Using silicon-on-insulator (SOI) technology, a thermal convection based inclinometer was reported in 2001, that can also be utilized as an accelerometer [32], [33]. SOI is a costlier variant of conventional Si wafer, with advantages like superior electrical insulation from the bulk substrate, and excellent etch stop and sacrificial layer functions due to the buried oxide layer underneath the active Si layer; hence, yielding better process control and device performance [33], [34]. The heater and thermistors of the reported device were made of lightly doped silicon, and the device was sealed in a TO-8 (Transistor Outline) metal can package with either air or $SF_6$ as the fluid in it. It was reported that the sensitivity increased with increase of package volume reaching maximum at a package volume of 12,000mm$^3$. With air as the working fluid, the sensitivity was 132µV/° and the response time was 110ms at heater power of 45mW. The sensitivity got enhanced to 6.6mV/° with $SF_6$ due to its higher density, though the response became slow (240ms). Another report on SOI-MEMS dual-axis accelerometer studied the effect of thermal stress (due to temperature changes) appearing in the Si thermistor structures causing its out-of-plane deformation and change in resistance due to piezoresistive effect, reducing the sensitivity [35]. The 2mm×2mm×0.4mm device as seen in figure 4, consists of four thermistors (of two different designs) arranged in a ring-like shape around the central heater (that was heated up to 200°C using 12.5mW). The novel shape allowed the thermistor to deform freely with temperature, hence, reducing the thermally induced stress by 90% in comparison to a usual clamped–clamped bridge thermistor. An off-chip circuit was employed to condition the sensor output. When tested in a measurement range of ±5g, the sensitivity was ~13mV/g (figure 4(c)) with a resolution of 10mg and the total noise (thermal and $I/f$) was estimated as 0.33µV.

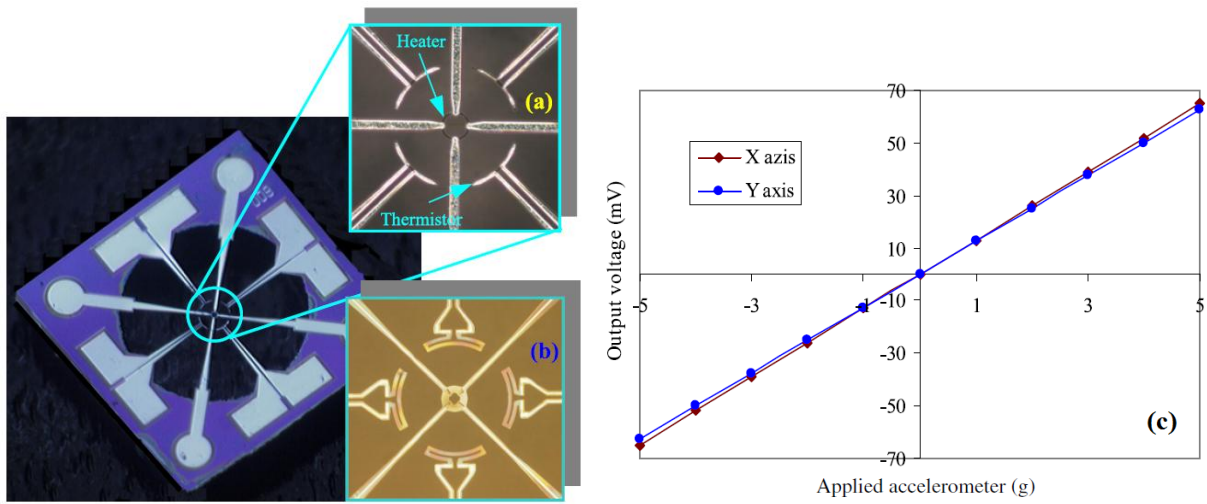

Figure 4. Micrograph of dual-axis thermal accelerometer with thermistors (of two designs (a) and (b)) arranged in a ring-like shape around the heater. (c) Experimentally measured sensor output voltage versus applied acceleration [35]. © 2007 IOP Publishing. (Reprinted with permission.)

The dependence of sensitivity on the cavity gas-pressure was studied experimentally in [36]. Here, platinum was used to realize the heater and thermistors, and the cavity dimension was 2mm×2mm×0.4mm. Pt while being expensive is preferred for its linear temperature coefficient (TCR) of resistance over a wide range, better reliability and accuracy of resistor, and higher thermal resistance with respect to other commonly used materials like Al or Si. However, Pt is not CMOS compatible as opposed to poly-Si, Al and Cu; so, can't be used in accelerometers in a CMOS process line. The device here was packaged in TO-16 metal can with the gas pressure being varied through a hole. The heater temperature was raised to 238K above ambient temperature by applying a heater power of 54mW. At atmospheric pressure, the sensitivity was 2.5mV/g. However, it increased proportionally to the square of gas pressure to a value of 138mV/g at 25 bar. Also, using three pairs of detectors placed at 100, 300 and 500μm from the heater, it was demonstrated that with an increase of gas pressure, the optimum sensitivity position comes closer to the heater.

Thermal accelerometers generally tend to become nonlinear in the case of a large characteristic length (i.e., large cavity size), at elevated levels of the heater temperature, and with higher acceleration. The linearity of convective accelerometers was studied by numerical and experimental methods in [37] using a device structure as in figure1. The simulation revealed that the device output was linear with acceleration for $Gr$ in the range of $10^{-2}$ to $10^{3}$. The sensitivity and linearity were both optimum when the temperature sensor was positioned at one-third distance between the heater and the cavity wall. The fabricated device had a cavity dimension of 3mm×2mm×0.25mm and the sensitivity was measured as 600μV/g in a range of 0 to 10g with an operating power of 87mW [38]. The bandwidth of the device was 75Hz. The NEA (due to thermal noise of the device) was 1 mg/√Hz at 25Hz. The resolution was found to increase (due to noise reduction) along with the sensitivity with increase in heating power. Although a large cavity volume ensures higher sensitivity due to increased heat exchange, but it leads the device towards nonlinearity region.

Garraud et al. [39] were able to measure high values of acceleration with good linearity by reducing the sensitivity of the device to 0.0045K/g. The sensitivity was reduced by means of reducing the cavity width to 600μm and by lowering the heater temperature. The device detected acceleration in the linear region up to 10,000g.

Sensitivity of a thermal accelerometer depends on the size and shape of the heater structure. Four types of heater structures were compared (in [40], [16]) and it was shown that a diamond-shaped heater (shown in figure 5) provided higher temperature gradient at the temperature sensing point in comparison to a square-shaped heater. The diamond-shaped heater was further modified by using high resistive material at the corners due to which the temperature gradient at the sensor position got increased. Al/poly-Si thermopiles were utilized here as temperature sensors. Unlike other commonly reported structures, here, back-side etching was also used in addition to front-side etching to create a well-defined cavity, minimizing performance variation due to cavity etching defects and ensuring device reliability and reproducibility. The sensitivity and bandwidth were 3.5mV/g and 25Hz respectively, at input power of 7.4mW and $SF_6$ as the working fluid. However, the sensitivity reduced to 30 µV/g with air. An illustration of the involved fabrications steps are also shown in the figure.

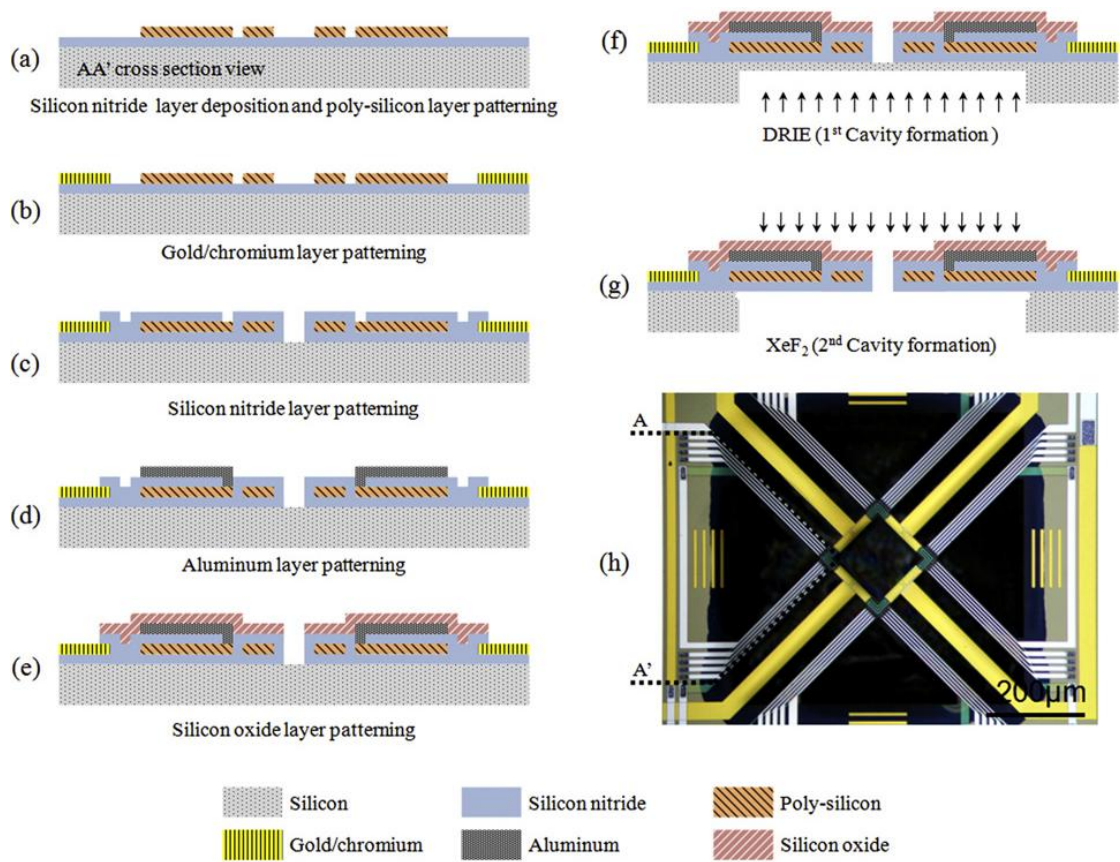

Figure 5. Fabrication process steps and an optical image of a micromachined dual-axis thermal accelerometer with diamond-shaped heater [16]. © 2011 Elsevier. (Reprinted with permission.)

To improve the bandwidth of thermal accelerometers, the cavity size should be made small and thermal diffusivity of the working fluid must be high. This was analyzed by Courteaud et al. [24] who obtained a -3dB bandwidth of 120Hz. The sensitivity of the device was 0.12K/g when the heater power and the characteristic dimension were 70mW and 1120µm respectively. It was also shown that the sensitivity decreased almost linearly with increase of the external ambient temperature. The effect of different working fluids like air, $CO_2$, $N_2$, water, Ar, He, and ethylene glycol in static and dynamic conditions were studied numerically in [41]. Different fluids yield different $Gr$ and $Pr$, hence, producing different heat convection behavior and temperature distribution within the accelerometer. This in turn affects the frequency response, sensitivity, and linearity of the device. The highest temperature difference hence, highest sensitivity was obtained using Ar due to its low dynamic viscosity. Air, $CO_2$ and $N_2$ produced practically identical results because of similar thermo-physical properties. The accelerometer's sensitivity and frequency response were experimentally studied as a function of the nature and pressure of fluid in [42] where a 320Hz bandwidth with He gas at 2.15bar was achieved. Bandwidth was found to increase proportionally with thermal diffusivity of the enclosed fluid, and decrease as the pressure of the fluid increases. Gases with lower molecular weight (like He) improves the frequency response, but increases the sensitivity to the ambient temperature of the package and hence, needs more heater power [43]. The use of inert gases is further preferred because other gases might cause the heater and detector to oxidize or age quickly. Liquids used as working fluid would produce a relatively slower response while requiring large heater power. An isopropanol-filled structure in [44] achieved a sensitivity of about 700 times of that of an air-filled accelerometer, but, having an order of magnitude greater response time.

Triple-axis accelerometers that can detect acceleration along all the three directions have considerably higher design and fabrication challenges. So, it took a while for such a convective device to be implemented. By means of integration of polymeric materials into MEMS process to attain the required mechanical flexibility, the first tri-axial accelerometer was reported in 2008 [45], [46]. Using surface micromachining on Si substrates with polyimide as a structural layer, out-of-plane/buckled sensing structures were assembled. A Cr/Au bilayer and a Ni layer were used to form thermocouple junctions for the thermopiles placed on the sensor plates. The measured sensitivity were 66, 64, and 25$\mu$V/g on the X, Y, and Z-axes respectively, using $SF_6$ as working fluid at a total heater power of 2.5mW. Later, a novel design consisting of three flexible polyimide membranes (two identical Z-axis membranes on either side of a central membrane) encapsulated within four polymeric microparts was reported, as seen in figure 6 [47], [48]. The central membrane included the heater and temperature sensors for X and Y-axes, while the upper and lower membranes had sensors for Z-axis. The X and Y sensitivity was about 8mV/g and the Z-axis sensitivity was 2.2mV/g, with an overall power of 45mW and the bandwidth was 4Hz.

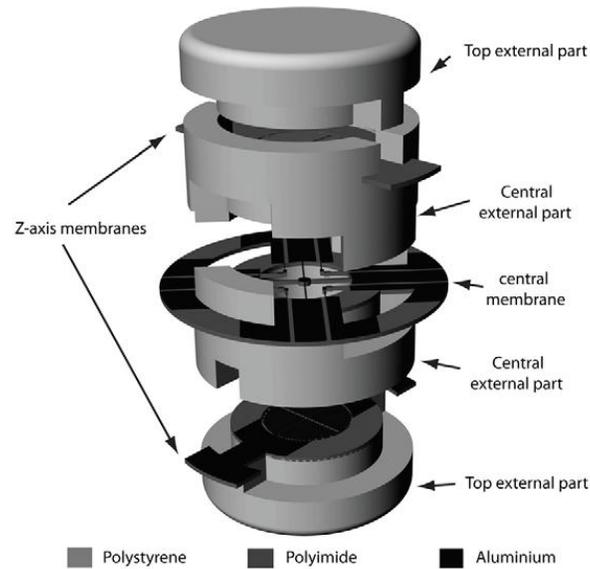

Figure 6. Tri-axial micromachined thermal accelerometer [47]. © 2011 Elsevier. (Reprinted with permission.)

## 4.2: Thermal Accelerometer in CMOS-MEMS Process:

In contrast to the accelerometers in the previous subsection, monolithic CMOS-MEMS thermal accelerometers can have the signal conditioning circuit on the same chip as the sensor as both are fabricated on the same substrate (die) using CMOS IC process in conjunction with some MEMS specific process steps [49], [50]. Hence, these are compact, cost effective, and insensitive to parasitic component effects. Moreover, the frequency response might also improve along with a lower power requirement. In the custom MEMS process based accelerometers discussed previously, various materials can be utilized for implementing the heater and the temperature sensor; and the material layer thicknesses can also be set as desired. But, in a standard CMOS process, the layer materials and thicknesses are predefined by the IC fabrication foundry. Moreover, the cavity size of the device cannot be made large as the device cost increases with the die size. Hence, the major limitation of thermal accelerometers implemented in this process is its comparatively lower sensitivity due to limited cavity size and available materials.

The foremost report was a single-axis convective accelerometer implemented in a 2μm CMOS process and packaged in ambient air by Milanovic et al. in 1998 [51], [52]. Both thermopile and thermistor types of temperature detectors were tested which yielded sensitivities of 136μV/g and 146μV/g respectively, with relatively lower power requirement (81mW) with thermopile. Good linearity in the range of 0 to 7g was observed for both the devices, and frequency response of up to hundreds of Hz was obtained. Sensitivity of the devices was found to be a nearly linear function of heater power (temperature).

A single-axis thermal accelerometer fabricated in AMS 0.8μm CMOS process is shown in figure 7 [25], [53]. The cavity was generated by front-side bulk micromachining process, and poly-Si resistors were used to realize the heater (40μm×1040μm) and thermistor (30μm×700μm) bridges having a separation of 200μm. The sensor output voltage was processed by a CMOS signal conditioning amplifier with controllable gain, which was present on the same chip as the sensor as seen in figure 7. The heater temperature was $438^{\circ}$C using 35mW of power and the device sensitivity was experimentally measured as 375mV/g (or equivalently, $1.53^{\circ}$C/g) with a resolution of 30 mg. It had good linearity till 10g and the -3dB bandwidth was 14.5Hz. Optimum device dimensions were formulated by means of FEM study of the effect of the cavity and package width and height on the device sensitivity and time constant [54], [55]. The sensitivity was found to increase linearly with the package cover height for a wide range. Due to etching defects, the cavity depth can get reduced due to which the thermal bubble size gets reduced leading to deterioration of sensitivity [55]. The temperature distribution and maximum sensitivity positions were obtained via 2D and 3D simulations and were compared in [56]. With respect to 2D, in 3D simulation, the thermal bubble got constricted and sensitivity reduced to a value closer to that obtained experimentally. The optimum sensitivity position for the temperature sensors also changed from middle to one-third distance between heater and cavity wall.

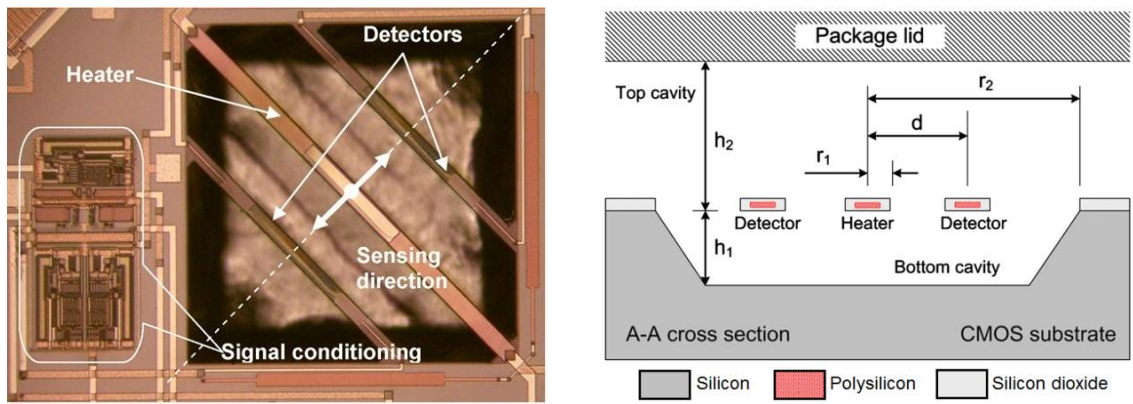

Figure 7. Prototype micrograph and schematic cross-section of thermal accelerometer fabricated in CMOS process with on-chip signal conditioner [53]. © 2008 Elsevier. (Reprinted with permission.)

A dual-axis thermal accelerometer fabricated in TSMC 0.35µm CMOS process was designed using a micro heater and two pairs of thermopiles being used as temperature sensors [57]–[59]. The cavity dimension was 830µm×830µm. The thermocouple was made by n-type poly-Si and Al junctions. The heater and thermopiles were connected by net like structures of $SiO_2/Si_3N_4$ to enhance the mechanical stability. The cold junctions were placed on the Si substrate and the hot junctions were formed over the net near to the heater. The device was tested as an inclinometer from which the acceleration sensitivity was derived to be 22µV/g at input power of 9.05mW. As the temperature is sensed at the junction of two different materials in a thermocouple, it requires less space at the hot junction where the temperature difference is measured. The noise voltage was under 0.25µV and the NEA was 0.159g.

Another reported dual-axis accelerometer fabricated in AMS 0.35µm CMOS process used meander-shaped heater of size 100µm×100µm and a cavity dimension of 600µm×600µm [14], [60]. The sensitivity was found to be 0.024K/g which is somewhat low due to a low Seebeck coefficient of 6.54µV/K of the Al/poly-Si thermopiles used, and lower cavity size and heater temperature. Acceleration was measured till 150g using heater temperature of $200^{\circ}$C. They also utilized infrared temperature mapping to experimentally determine the temperature profile within the device. An optimized square-ring shaped heater structure was reported in [61] to improve the sensitivity. It was also shown that the sensitivity improved when a given peak temperature is generated at the edges of the heater rather than at its center.

A monolithic triple-axis thermal accelerometer was implemented by Mailly et al. using AMS 0.35µm CMOS process, which didn't require the complex assembly operations of the tri-axial sensors as discussed in the previous subsection [62], [63]. The principle is to sense the acceleration in the third axis by means of common-mode temperature measurement using X and Y detectors of a 2-axis convective accelerometer structure as illustrated in figure 8. Acceleration in the positive Z direction stretches the hot bubble leading to temperature drop at the detectors, while reverse occurs due to negative Z-axis acceleration. Experimentally obtained sensitivity values were 8.8mV/g, 12.6mV/g and 0.45mV/g for X, Y and Z-axes respectively at heater power of 8.3mW. In comparison to the in-plane sensitivity, the low Z-axis sensitivity was attributed to unoptimized thermal sensor position and 3D thermal effects that demand proper understanding. Accelerometers based on a similar sensing method have also been commercialized by MEMSIC [64], [65].

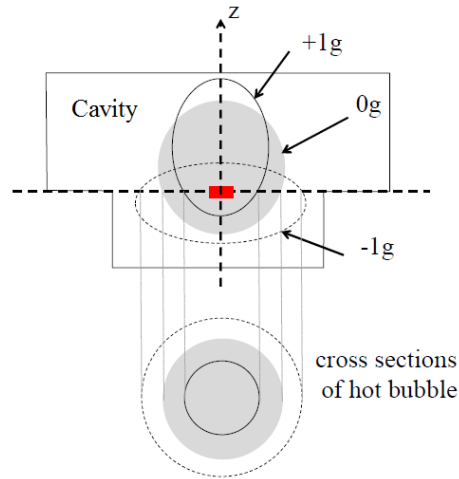

Figure 8. Detection of out-of-plane acceleration using in-plane temperature detectors within a cavity that is asymmetric along the Z-axis [63]. © 2014 Springer. (Reprinted with permission.)

## 4.3: New and Innovative Devices:

A number of unique instances of reported thermal accelerometers that incorporate innovative materials and structures have been presented in this subsection. Such modifications are targeted towards enhancement of the different performance aspects.

### 4.3.1 Organic & Plastic Substrate

Petropoulos et al. [66] reported a thermal accelerometer fabricated on an organic PCB (printed circuit board) substrate with heater and temperature sensors made of platinum. The substrate had a much lower thermal conductivity ($0.2Wm^{-1}K^{-1}$) than that of conventionally used silicon or $SiO_2$, hence, providing better insulation for leakage of the heater power through the substrate. The working fluid used was water which was covered by a tank of dimension 3.5cm×1.5cm and a depth of 850μm. When 60mA current was passed through the heater, the sensitivity obtained from the device was 32mV/g. The achieved sensitivity was also high due to large size of the sensor and higher operating power. On a similar note, a flexible polyimide substrate with low thermal conductivity was used in [67]. Further, they compared $CO_2$ and xenon gases as working fluid and concluded that the sensitivity was larger with $CO_2$; but with Xe, higher levels of acceleration (25g) could be measured without saturating and the response was also faster.

### 4.3.2 CNT Based Components

The heater and temperature detectors were fabricated using multi-walled carbon-nanotubes (MWCNT) by Zhang et al. on 1mm-thick glass substrate [68], [69]. The advantage of using

CNTs in the device is a ultra-low power requirement (in the order of tens of pW) and smaller size of the detectors. MWCNT has negative TCR, which means a drop in the output voltage of the thermistors with increase in its temperature. The accelerometer's response was tested at three ranges of the heater power. At substantially low value of the heating current (< several nA), the detector temperature gets reduced towards ambient temperature and measurement noise became dominant. On the contrary, when the heating current was very high (several hundred nA), the sensor response became too low due to heating of the CNT as well as the connecting electrodes, surrounding air and the substrate. This is because the acceleration induced convection couldn't quite affect the temperature of the CNTs which showed very low resistance change. For current levels in between these two extremes (~0.1μA), the device worked properly and produced a linear-log relationship between the sensor response and applied acceleration. The sensor was specifically found to be sensitive in detecting small acceleration values (~0.1m/s$^2$) with the response getting saturated at higher acceleration.

A novel thermal convective inclinometer using yarn of CNT for both the heater and thermal detectors was reported recently that also consumes significantly smaller heater power of 33μW and yields a sensitivity of 1.8μV/g [70]. The CNT yarn exhibits a stable resistance over a wide range of temperatures, hence producing good device linearity.

### 4.3.3 Porous Silicon Based Device

Instead of having front-side bulk micromachining to make a cavity, a 60μm thick porous silicon (PS) layer was used for thermal isolation between the thermal accelerometer heater and the underlying substrate in [71]. The PS layer has a thermal conductivity of 1.2 W/mK that helps to confine the generated heat within the desired region by thermally isolating the Si substrate. So, the device can be operated at relatively low power. Also, due to the elimination of any freestanding structures here, the sensor becomes more immune to shock, aging effects, and calibration errors. Its heater was designed using poly-Si and the temperature sensor using poly-Si /Al thermopiles. The die size was 1.4mm×0.9mm and the device was packaged in ceramic DIP-8. The reported device sensitivity was 13mV/g, linearity range was up to 6g acceleration, and the bandwidth was around 12–70Hz. The power supplied to the heater was 166mW. The device was also tested using different packaging of air and oil (SAE 20) as working fluid, namely, with non-sealed air, non-sealed oil, and sealed oil package [72], [73]. Due to much higher viscosity and thermal conductivity of oil with respect to air, a higher sensitivity was found in the latter two cases; the highest being in the second case due to free oil surface. However, the best linearity was achieved by the third configuration due to the absence of fluid surface movement.

### 4.3.4 Novel Device Structures

In a recent report, the cavity structure of accelerometer was modified to improve the sensitivity [74]. Silicon islands were placed in between the heater and temperature sensors as in figure 9. Since the thermal conductivity of Si is much more than that of the fluid, presence of Si islands modulates the fluid movement and hence, the temperature profile. The sensitivity obtained from the cavity with island structure was 0.657K/g which is double of that without islands (0.335K/g).

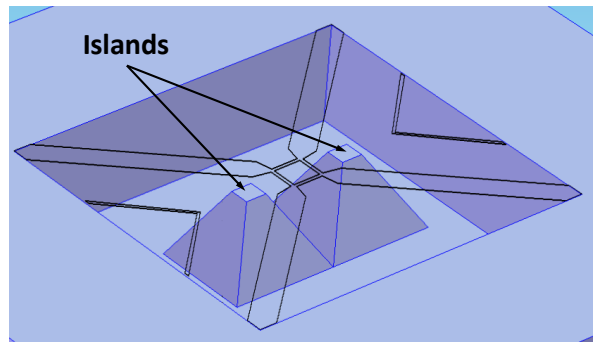

Figure 9. Silicon islands within the cavity enhance the sensitivity of the convective accelerometer.

Due to the good temperature sensitivity (typically −2.2mV/°C) of a diode [75], it can be utilized as the temperature sensor in thermal accelerometer to improve its sensitivity. Additionally, the sensitivity can be further enhanced by connecting two or more diodes in series [76]. However, it is extremely challenging to integrate Si diode in the cavity because of the need to etch silicon to realize the suspended membrane within the cavity.

### 4.4: Commercial Thermal Accelerometers:

Capacitive sensing being the most common and popular accelerometer variety in the market, it seems pertinent that its performance aspects be compared with that of a thermal accelerometer. While a large number of manufacturers like Analog Devices, Bosch, NXP/Freescale, STM, TDK/InvenSense, etc. market capacitive MEMS acceleration sensor chips, MEMSIC is currently the only major player providing commercially available thermal accelerometers since 2002. In table 2, a tri-axial thermal accelerometer from MEMSIC manufactured in standard submicron

CMOS process with Al/poly-Si thermopiles [77] have been compared against an Analog Devices capacitive sensor that uses a poly-Si surface-micromachined proof-mass built on top of Si wafer [78]. The devices have been selected to have the same DOF, comparable measurement range, and both provides analog output voltages that are proportional to the applied acceleration. As evident, the thermal one has a remarkable shock survival rating while its frequency response is comparatively poor.

**Table 2.** Comparison of commercially available thermal and capacitive accelerometer types.

|  | **Thermal accelerometer** | **Capacitive accelerometer** |
|---|---|---|
| Model number | MXR9500MZ | ADXL327 |
| Sensitivity | 500mV/g | 420mV/g |
| Bandwidth | 17 Hz | 550/1600 Hz |
| Full scale range | ±1.5g | ±2g |
| Nonlinearity | 0.5% of full scale | ±0.2% of full scale |
| RMS noise density | 0.6/0.9 mg/√Hz | 0.25 mg/√Hz |
| Cross-axis sensitivity | ±2% | ±1% |
| Mechanical shock survival | 50,000g | 10,000g |
| Supply current | 4.2mA at 3.0V | 350µA at 3.0V |
| Package dimension | 7mm × 7mm × 1.8mm | 4mm × 4mm × 1.45mm |

A latest offering MXC4005XC from MEMSIC uses a relatively new wafer-level packaging technology that integrates the step of packaging with the fabrication of the wafer which is later diced yielding packaged chips that are practically of the same size as the die. This is very efficient in terms of size and cost, with a reported shock survival of 200,000g.

**5. Properties of Materials Used in Accelerometer:** Different materials used for manufacturing of convective inertial sensors can be found in literature, with each offering distinctive benefits and drawbacks. Proper understanding of the various properties of these materials is imperative for successful realization of the desirable device performance. For example, a material like poly-Si having quite low electrical conductivity requires much less area for making a heater or thermistor, but its TCR can be an issue. For Al/poly-Si thermocouples, the Seebeck coefficient is an important factor that depends on the doping. If the Seebeck coefficient is high, the voltage change per °C will be better providing higher sensitivity. Regarding the working fluid present in the cavity, as discussed previously, properties like $Gr$ and $Pr$ are important for performance like sensitivity and response time of the accelerometer. In order to reduce the power requirement, a material with low thermal conductivity is appropriate as the substrate to prevent the outflow of generated heat energy through it. Table 3 summarizes some relevant thermophysical and electrical properties of materials commonly used in implementing thermal accelerometers.

**Table 3.** Properties (at ~300K) of some common materials used in thermal accelerometer fabrication [33], [79]–[84].

| Material | Thermal conductivity $k$ (Wm$^{-1}$K$^{-1}$) | Electrical conductivity $\sigma$ ($\Omega^{-1}$m$^{-1}$) | Density $\rho$ (kg m$^{-3}$) | Dynamic viscosity $\mu$ (Pa s) | Specific heat $C_p$ (J kg$^{-1}$ K$^{-1}$) | Thermal diffusivity $\alpha$ (m$^2$s$^{-1}$) |
|---|---|---|---|---|---|---|
| Silicon | 156 | $2.33 \times 10^{-4}$ [a] | 2,330 | — | 707 | $97.52 \times 10^{-6}$ |
| Polysilicon | 31 | $2.6 \times 10^{-3}$ [b] | 2,330 | — | 707 | $16.5 \times 10^{-6}$ |
| Silicon dioxide | 1.4 | $3 \times 10^{-13}$ | 2,270 | — | 1000 | $6.2 \times 10^{-7}$ |
| Platinum | 71.6 | $9.4 \times 10^6$ | 21,500 | — | 133 | $24 \times 10^{-6}$ |
| Aluminum | 235 | $3.7 \times 10^7$ | 2,710 | — | 904 | $93 \times 10^{-6}$ |
| MWCNT | 750 [c] | $10^4$–$10^7$ | 1650 | — | 730 | $4.6 \times 10^{-4}$ |
| Air | 0.026 | — | 1.2 | $1.9 \times 10^{-5}$ | 1005 | $22 \times 10^{-6}$ |
| Carbon dioxide | 0.017 | — | 1.7 | $1.5 \times 10^{-5}$ | 850 | $1.1 \times 10^{-5}$ |
| Argon | 0.018 | — | 1.6 | $2.27 \times 10^{-5}$ | 521 | $21 \times 10^{-6}$ |
| SF$_6$ | 0.013 | — | 6.14 | $1.6 \times 10^{-5}$ | 598 | $3.5 \times 10^{-6}$ |
| Water | 0.6 | — | 997 | $8.3 \times 10^{-4}$ | 4071 | $1.52 \times 10^{-7}$ |
| Oil | 0.145 | — | 900 | 0.3 | 1910 | $8.5 \times 10^{-8}$ |

[a] Depends on doping. [b] Depends on grain boundary structure and doping. [c] Effective value is smaller due to coupling within MWCNT bundles, sheet imperfections, etc.

**6. Signal Conditioning for Thermal Accelerometers:** The temperature gradient generated within a thermal convection accelerometer produces a corresponding resistance variation of its thermistor temperature detectors due to their positive temperature coefficient of resistance. As illustrated in figure 10, the thermistors can be arranged in a Wheatstone bridge along with a pair of reference resistors which are present on the substrate and hence, maintained at ambient temperature [25], [53]. Thus, the bridge generates a differential output voltage proportional to the applied acceleration. If a pair of thermopiles are instead used as temperature sensors, a differential voltage signal is directly produced [57], [14]. But, the electrical output being very small in magnitude, is susceptible to noise. Thus, like most other sensors, in order to ensure an accurate measurement, the analog output signal requires suitable conditioning before it can be provided to an analog-to-digital converter (ADC) which enables further digital processing and acquisition of the signal [85], [86]. For thermal inertial sensors, the signal conditioning requirements mainly encompass amplification and filtering. One or multiple amplifiers are necessary for boosting the level of the analog signal to match the full dynamic range of the succeeding ADC, while ensuring minimal noise addition, low offset, and a proper impedance match with the sensor. Filters might be needed to remove noise at frequencies outside the signal of interest, and also for anti-aliasing requirement before the signal is sampled by the ADC. Further, as the signal to be acquired has a low frequency range (dc to ~200Hz), hence, data converters like successive approximation register (SAR) and sigma-delta ($\Sigma\Delta$) ADCs are appropriate for thermal acceleration sensor applications.

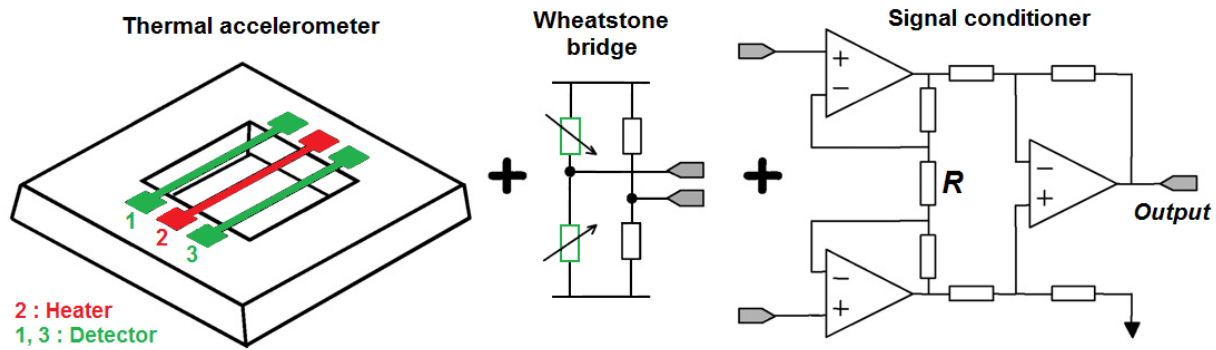

Figure 10. Thermal acceleration sensor and signal readout electronics [25]. © 2006 Elsevier.
(Reprinted with permission.)

Due to low frequency of the accelerometer signal, the effect of Flicker (1/f) noise becomes significant. Also, as in any signal conditioning chain, the noise contribution from the first circuit block determines the noise performance of the overall system, so, the first amplifier should preferably be a low-noise instrumentation amplifier (IA). For their readout interface, Chaehoi et al. [25] have used an on-chip instrumentation amplifier implemented with a popular three operational amplifier topology, having programmable gain (of 10, 100 or 1000) by means of controlling the resistor R (figure 9). Circuit techniques for minimizing 1/f noise of the IA, like chopper stabilization, correlated double sampling, large sized p-type MOS transistors for the differential input pair, etc. can be applied [53], [87]. Further, instead of measuring voltage, the output current from the thermopile detectors can be sensed using a low-noise transimpedance amplifier as demonstrated by Goustouridis et al. [73]. Process variations in the temperature sensing resistors might lead to significant values of dc-offset at the input of the instrumentation amplifier causing the output of the signal conditioner to get saturated. To avoid this, the signal chain may be provided with the provision of cancelling out the dc-offset presented to its input without affecting any useful input dc signal [88].

A custom designed signal conditioning analog front-end can be used (integrated monolithically as in figure 6, or at the package-level); or, a suitable commercially available off-chip conditioning IC may be utilized for the thermal acceleration sensor [46], [49]. Of course, the former *system-on-chip* solution is preferable as the circuit can be tuned according to sensor specific needs; and also due to the aspects of assembly and packaging size, cost, and the effects of parasitics from interconnects and bond-pads that might impede the system's performance. HDL (hardware description language, e.g., Verilog-A) models of convective accelerometer have been developed that enables its co-simulation with associated CMOS readout electronics using SPICE circuit simulators [53]. This provides opportunity to optimize the designs of the sensor and the circuit in synchronization in order to improve the overall system performance.

Innovative *closed-loop* convective accelerometers have been reported where the inertial sensor has been placed within a negative thermal feedback loop. This decreases the thermal response time, consequently yielding an increased bandwidth, without reduction in sensitivity [87], [89]. Garraud et al. implemented a closed-loop design by adding two resistors placed close to the temperature detectors as depicted in figure 11 [90]. The bias currents of these resistors were appropriately altered (the feedback signal) using an electronic PID controller, so as to rebalance any temperature difference produced between the two detectors from applied acceleration. By this, a closed-loop bandwidth of 1025Hz was attained using an usual thermal accelerometer of ~70Hz bandwidth. Another reported closed loop implementation is a thermal *sigma-delta modulator* having the thermo-electrical (first-order low-pass) response of the convective accelerometer as its loop-filter [91], [92]. This again provides an improved bandwidth, and also a direct digital output. Frequency response compensation circuit to extend the frequency response of thermal accelerometer was provided in [43]. As the sensitivity of thermal accelerometer depends on temperature dependant properties like density, specific heat, dynamic viscosity and thermal conductivity of the fluid within its cavity, circuitry to compensate the variations of sensitivity over a range of temperature and supply voltage can also be provided on-chip [93]. Further, self-test circuit for continuous checking of the integrity of the heater, detector and associated circuitry of the accelerometer can also be made available to increase the device reliability [93], [94]. Lin and Lin [67] achieved to exhibit a wireless thermal accelerometer by having an RFID (radio-frequency identification) tag flip-chip bonded to it.

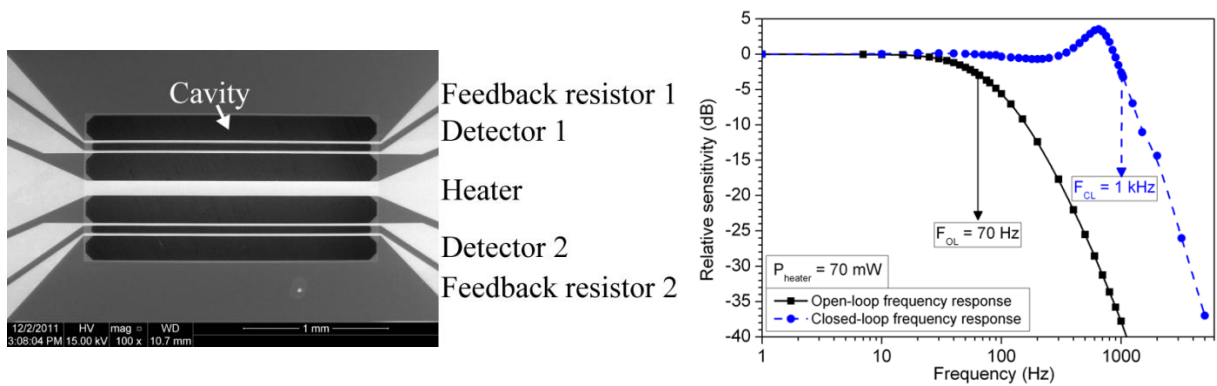

Figure 11. Scanning electron microscope image and measured frequency response of convective accelerometer with thermal feedback arrangement [90]. © 2012 IET. (Reprinted with permission.)

**7. Applications of Thermal Accelerometer:** Accelerometers in general are used for a wide range of sensing applications by automotive, avionics, military, robotics, and consumer goods industries. Some of the potential application areas for thermal accelerometers are as follows:

i. Automotive applications: computerized electronic stability control improves a vehicle's stability by detecting loss of traction, skidding, rollover, etc. and automatically regulating the engine throttle and braking. Convective inertial sensors are well suited for use in such systems [95]. Further, usage in car air-bags that are meant to protect the passengers by instantaneously inflating in the scenario of a vehicular collision is also feasible [43].

ii. Predictive drop sensor: In a computer hard-disk, the read/write header can get damaged due to a free fall. A thermal accelerometer provided inside the disk can be used to detect the free fall and immediately trigger a mechanism by which the disk's header can be safely parked preventing a crash [96].

iii. Tilt/angle detection: Along with acceleration, convective inertial sensors can also be used as gyroscopes to detect orientation and rotation of devices. These are required in consumer electronics such as digital camera, virtual-reality device, gesture recognition, and joystick [32], [33], [70].

iv. Vibration detection: Constant vibration within any machinery creates wear and tear on its parts like the bearing, seals and couplings. Machine-mounted inertial sensors can be used to determine the condition of the machine as well as to predict the precise cause and location of problems before any considerable damage might occur [97]. This may also be used for structural health monitoring of bridges, buildings, and aerospace systems [98].

v. Steady threshold switch: One of the advantages of thermal inertial sensor is its high shock survival rating. Using a shocking hammer capable of generating an acceleration of up to 50,000g, and commercially available dual-axis accelerometer, the effect of heavy shock was tested [99]. With such a large impact, the output of the convective accelerometer got saturated. It was concluded that such devices could be used to determine if the applied impact has crossed a certain threshold level.

**8. Outlook and Conclusions:** The rapid progress of MEMS research, manufacturing, and marketing in the past few years coupled with the increasing demand for inertial sensors strengthen the outlook for thermal convective sensors. This article is intended as a guide for the researchers in this domain providing a comprehensive review of the theory, modeling, simulation, research innovations, involved materials, applications, and critical performance aspects. Few instances of convective accelerometers are already being aggressively marketed. In spite of the advantages, thermal inertial sensors in their present form are not competitive enough in the key aspects of acceleration sensitivity and frequency response. As discussed, prior efforts made to improve the sensitivity and bandwidth include modifying the cavity, heater, and package geometries, using different working fluids, and optimizing the sensor position. A larger cavity size may improve sensitivity while sacrificing the bandwidth, but, will require larger area for fabrication increasing the cost and deteriorating the mechanical stability. Usage of some fluids to gain sensitivity might complicate the packaging. Intelligently designed closed-loop systems and frequency response compensation circuits are excellent method of enhancing the bandwidth and needs to be explored further. Improving the out-of-plane sensitivity of triple-axis convective sensors is also needed. Thus, sensitivity and bandwidth improvement remains an area of active research that needs to be addressed in the near future. With regard to power consumption, the main contributors are the heater, and the leakage through the substrate. Higher heater power enhances its temperature that helps to increase the sensitivity. To reduce the power, different innovative heater, thermistor, and substrate structures and materials have been experimented with in literature. But, low energy requirement being a fundamental necessity for modern battery-operated systems, keen focus needs to be maintained in lowering the power. Further, design of custom temperature compensation and self-test/calibration circuitry for enhancing the device reliability is an important research frontier. Investigations may also be done towards amalgamating the concept of thermal accelerometer with other innovative inertial sensing solutions to improve the overall performance, like with a recently proposed liquid state accelerometer that employs a tiny electrolyte droplet as the sensing body over four electrodes (anode-cathode-cathode-anode) used for read-out [100]. With applied acceleration, the droplet moves and the electrochemically induced output current (between the anode-cathode pairs) changes due to convective transport of ions between the electrodes. Thermal convective sensors are just over two decades old, and hence, are expected to witness much more research attention in the near future revealing exciting developments. Its inherent features of superior shock survival, simplistic compact structure, low cost, wide measurement range, and integrability with CMOS will certainly help in garnering the necessary attention.